\documentclass[prl,aps,twocolumn,showpacs]{revtex4}

\usepackage{epsfig}
\usepackage{graphicx}
\newcommand{\ket}[1]{|#1\rangle}

\begin{document}

\title{Optimal Topological Test for Degeneracies of Real Hamiltonians}

\author{Niklas Johansson and Erik Sj\"oqvist\footnote{Electronic 
address: eriks@kvac.uu.se}}

\affiliation{Department of Quantum Chemistry, 
Uppsala University, Box 518, S-751 20 Sweden}

\begin{abstract} 
We consider adiabatic transport of eigenstates of real Hamiltonians 
around loops in parameter space. It is demonstrated that loops that
map to nontrivial loops in the space of eigenbases must encircle
degeneracies. Examples from Jahn-Teller theory are presented to
illustrate the test. We show furthermore that the proposed test is
optimal.
\end{abstract}
\maketitle
Sign reversal of real electronic eigenfunctions when continuously
transported around a degeneracy was discovered by Herzberg and
Longuet-Higgins \cite{H_LH63} and was later used to construct a
topological test for conical intersections \cite{L_H75}. 
Such intersections are abundant in molecular systems \cite{Tru_Mea03}
and are important because they signal a breakdown of the
Born-Oppenheimer approximation. The topological test in 
\cite{L_H75} has been used to detect conical intersections 
in LiNaK and ozone \cite{Testusers}. The sign change of the 
electronic eigenfunctions gives rise to the molecular Aharonov-Bohm 
effect \cite{MAB}, which has recently been experimentally tested 
\cite{Busch98} and theoretically investigated \cite{Sjo02_Ken97}. 
The structure of adiabatic wave functions and the sign change pattern
in the vicinity of degeneracies have also been analyzed in the context
of quantum billiards. The behavior of the real wave functions for such
systems, studied by analog experiments on microwave resonators
\cite{Lauber94}, has been interpreted in terms of both the standard
\cite{Man_Ch99} and the off-diagonal \cite{Pist_Man00} geometric
phases. The microwave resonator experiments have motivated
general theoretical treatments concerning both the concomitant 
geometric phases and structure of the wave functions
\cite{Man_Ch99, Pist_Man00, Sam_Dhar0102}.

In this Letter we study the eigenvectors of a real matrix
Hamiltonian on a loop in parameter space. We show that the behavior of
the eigenvectors may imply the presence of a degeneracy even if none
of the eigenvectors changes sign around the loop. This result is a
generalization of Longuet-Higgins' topological test for intersections
\cite{L_H75}. It is also proven that the generalized test exhausts all
topological information associated with the behavior of the
eigenvectors, concerning the presence of degeneracies.

We begin by the following topological fact. Let $X$ and $Y$ be 
topological spaces. If $\Gamma$ is a trivial loop in $X$, and if 
$F:X \rightarrow Y$ is continuous, then $F(\Gamma)$ is a trivial 
loop in Y.

To prove this, note that if $G$ is a homotopy between $\Gamma$ and a
point $x_{0} \in X$, then $F\circ G$ is a homotopy between $F(\Gamma)$
and $F(x_{0}) \in Y$.
 
Now let $H(Q)$ be an $n\times n$ matrix Hamiltonian, written in the
basis $\{|i\rangle\}$ of the $n$ dimensional Hilbert space ${\cal
H}$. We suppose that $H(Q)$ is real, symmetric, and continuous for
each $Q = (Q_{1}, \ldots,Q_{d})$ in parameter space ${\cal Q }$, which
we assume to be a simply connected subset of ${\bf R }^{d}$. The
eigenvectors of $H(Q)$ can always be chosen real. We call the space of
real vectors ${\cal H}_{real}$, which through the expansion
coefficients in the basis $\{|i\rangle\}$ can be identified with
${\bf R}^{n}$. The set ${\cal N}_{real}$ of normalized real vectors 
is the sphere $S^{n-1}$ in ${\bf R}^n$.

Consider a simply connected surface $S$ in ${\cal Q}$, bounded by the
loop $\Gamma$. Longuet-Higgins' theorem asserts that if a certain
eigenvector $|\psi_{i}(Q)\rangle$ of $H(Q)$ changes sign when
continuously transported around $\Gamma$, then there is a point on $S$
where $|\psi_{i}(Q)\rangle$ becomes degenerate with another state. We
note that the theorem implies that if $H(Q)$ is nondegenerate on $S$,
then $\pm |\psi_{i}(Q)\rangle$ represent two continuous functions from
$S$ to $S^{n-1}$.

As a first step towards a generalization of Longuet-Higgins' theorem
we consider two-level systems. The set of normalized vectors ${\cal
N}_{real}$ is then the circle $S^{1}$. We prove that if an eigenvector
of a real two-level system represents a nontrivial loop in $S^{1}$ 
when continuously transported along a loop $\Gamma$, then $\Gamma$ 
must encircle a degeneracy. The proof is by {\it reductio ad
absurdum}. Suppose that $H(Q)$ is nondegenerate on the surface $S$
bounded by $\Gamma$. A real eigenvector $|\pm (Q)\rangle$ then
represents a continuous function $F$ from $S$ to ${\cal N}_{real} =
S^{1}$. Since $\Gamma$ is trivial ($S$ being simply connected), so 
is the loop traced out by $|\pm (Q)\rangle$.

As an illustration, let us consider the coupling matrix Hamiltonian of
the $E\otimes \epsilon$ Jahn-Teller system \cite{Z_G87}
\begin{eqnarray}
 & & H(\rho,\theta)
\nonumber \\ 
 & = & \left( \begin{array}{cccc} 
k\rho \cos \theta + \frac{1}{2} g\rho^{2} \cos 2\theta& 
k\rho \sin \theta - \frac{1}{2} g\rho^{2} \sin 2\theta\\
k\rho \sin \theta - \frac{1}{2} g\rho^{2} \sin 2\theta &
-k\rho \cos \theta - \frac{1}{2} g\rho^{2} \cos 2\theta
\end{array} \right), 
\nonumber \\ 
\label{Exe}
\end{eqnarray}
where $\rho$ and $\theta$ are polar coordinates of parameter space
${\cal Q}={\bf R}^{2}$, and $k$ and $g$ are the linear and quadratic 
coupling strengths, respectively. In this system there are four 
degeneracies: one at the origin $\rho = 0$ and three at $\rho = 2k/g$ 
and $\theta = \pi/3$, $\pi$, and $5\pi/3$. Continuous transport of 
an eigenvector around any single degeneracy produces a sign change 
\cite{Z_G87}. Longuet-Higgins' test applied to such a loop would 
thus imply a degeneracy.
Consider instead a circular loop $\Gamma$ given by $\rho \gg 2k/g$ 
and $\theta \in [0,2\pi]$, encircling all four degeneracies. The 
eigenvectors $|-(\theta)\rangle \approx
\sin\theta|1\rangle + \cos\theta|2\rangle$ and $|+(\theta)\rangle 
\approx -\cos\theta|1\rangle + \sin\theta|2\rangle$ do not change 
sign as the loop is traversed. However, viewed as elements of the 
circle $S^{1}$, $|\pm(\theta)\rangle$ both make a complete clock-wise 
turn. This means that each eigenvector traces out a nontrivial 
loop in ${\cal N}_{real} = S^{1}$. In this way the presence of 
degeneracies encircled by $\Gamma$ can be detected.

For the two-level case we have thus obtained a generalization of
Longuet-Higgins' test: also loops along which the eigenvectors trace
out nontrivial loops in $S^{1}$ encircle degeneracies on every
surface bounded by them. However, for $n\geq 3$, the space of real
normalized vectors ${\cal N}_{real} = S^{n-1}$ is simply connected,
i.e., it contains only trivial loops. Thus, following a single
eigenvector along the loop is insufficient. To generalize the test in
this case, we consider instead a complete set $\{|\psi_{i}(Q)\rangle
\}_{i=1}^{n}$ of normalized eigenvectors of $H(Q)$.

If $H(Q)$ is nondegenerate on $S$, then $\pm |\psi_{i}(Q)\rangle$
are, for each $i$, two continuous and globally defined
functions from $S$ to $S^{n-1}$. Without loss of generality we 
may assume that $\{|\psi_{i}(Q)\rangle \}_{i=1}^{n}$ represents 
a positively oriented orthonormal basis of ${\cal H}_{real} = 
{\bf R}^{n}$. Every such basis can be thought of as an element 
of the $n$ dimensional rotation group SO($n$). We may thus define 
a continuous function $F:S\rightarrow \mbox{SO}(n)$ as
\begin{equation}\label{F}
F(Q) = \left( \begin{array}{ccc} 
\langle 1|\psi_{1}(Q)\rangle & \ldots & \langle 1|\psi_{n}(Q)\rangle \\ 
       \vdots                & \ddots &           \vdots \\
\langle n|\psi_{1}(Q)\rangle & \ldots & \langle n|\psi_{n}(Q)\rangle \\
\end{array} \right).
\end{equation}
Thus, if $H(Q)$ is nondegenerate on $S$, then its eigenvectors
represent a continuous function from $S$ to SO($n$). Analogous to the
two-level case it follows that the loop in SO($n$) traced out by
$F(Q)$ as $Q$ varies along $\Gamma$, is a trivial loop. By
\textit{reductio ad absurdum} we consequently arrive at the following
result. If the $n$ eigenvectors of $H(Q)$ represent a nontrivial loop
in SO($n$) when taken continuously around $\Gamma$, then there must be
at least one degeneracy of $H(Q)$ on every simply connected surface
$S$ bounded by $\Gamma$.

This result makes it possible to detect the presence of a degeneracy
by considering eigenvectors on a loop in ${\cal Q}$ even if they do not
change sign around the loop. It constitutes the promised
generalization of Longuet-Higgins' test.

We note that $\mbox{SO}(2)$ is homeomorphic to $S^{1}$ and is thus
infinitely connected: it contains one class of nontrivial loops for
each nonzero integer. The fundamental group of SO($2$) is the additive
group of integers ${\bf Z}$. Determining whether a loop is trivial
amounts to counting the number of times the eigenvector encircles the
origin. For each $n \geq 3$, however, SO($n$) contains only one class
of nontrivial loops, that all become trivial when traversed twice. The
fundamental group of SO($n\geq 3$) is the two-element group ${\bf
Z}_{2}$. In order to apply the test in this latter situation, we need
to know how to determine whether a loop in SO($n$) with $n\geq 3$ is
trivial or not. In principle, this can be done by lifting the loop to
the universal covering space Spin($n$) \cite{Jost} of SO($n$). In
practice, though, it seems difficult to find a method, that works for
all $n$ and is easily implemented. Below we show explicitly how to use
the test in the cases $n=3$ and $n=4$. We use examples from
Jahn-Teller theory as physical illustrations.

First, let us consider the $n=3$ case. SO($3$) is homeomorphic to the
closed ball of radius $\pi$ with antipodal points on its surface
identified \cite{tung}. A vector $\phi {\bf \hat{v}}$ in the ball
represents a rotation around the unit vector ${\bf \hat{v}}$ by the
angle $\phi \in [0,\pi]$. Antipodal points on the surface must be
identified since a rotation by the angle $\pi$ is the same
transformation regardless of whether the rotation axis is plus or
minus ${\bf \hat{v}}$. A loop in SO($3$) can thus be viewed as a curve
that may exit the closed ball at the boundary, and enter
again at the antipodal point. The loop is trivial if and only if the
number of piercings of the boundary is divisible by two. 
Such a piercing is characterized by $\phi = \pi$ and that 
${\bf \hat{v}}$ abruptly changes sign \cite{remark}. 

\begin{figure}[tbp]
\centering
\includegraphics[width=8cm]{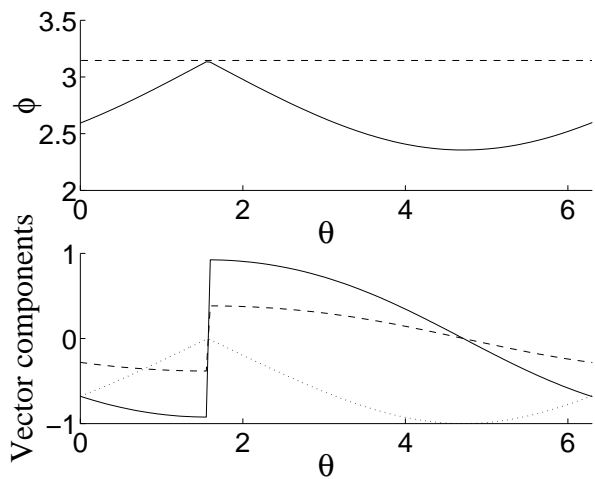}
\caption{Rotation angle $\phi$ in radians and components of the 
rotation vector ${\bf \hat{v}}$, as functions of $\theta$ also 
in radians. $\phi$ and ${\bf \hat{v}}$ are computed from the matrix 
$F(\theta)$ of Eq. (\ref{FTxt}) representing eigenvectors in the 
$T\otimes \tau_2$ system. The first, second, and third component of 
${\bf \hat{v}}$ are represented by the solid, dotted, and dashed 
curves, respectively. Note that $\phi$ equals $\pi$ at 
$\theta = \pi/2$, and that ${\bf \hat{v}}$ changes sign at that point.}
\label{components}
\end{figure}

We apply the method to the linear $T\otimes \tau_{2}$ Jahn-Teller
system, described by the coupling matrix \cite{OBrien89}
\begin{equation}\label{Txt}
H({\bf R}) = \left(\begin{array}{ccc}
 0 & -Z & -Y \\
-Z &  0 & -X \\
-Y & -X & 0  \\
\end{array} \right),
\end{equation}
where ${\bf R} = (X, Y, Z)$ parametrizes ${\cal Q} = {\bf R}^{3}$. The
Hamiltonian is doubly degenerate on eight rays in ${\bf R}^{3}$ that
go out from the origin and into the middle of each octant. At the
origin there is a three-fold degeneracy. We consider a loop $\Gamma$
lying completely in the $X$-$Y$ plane. In planar polar coordinates
$\rho$ and $\theta$, the eigenvectors arranged according to increasing 
energy read 
$|\psi_{1}(\theta)\rangle = \frac{1}{\sqrt{2}} \big( \sin\theta 
|1\rangle + \cos\theta |2\rangle + |3\rangle \big)$, 
$|\psi_{2}(\theta)\rangle =\cos\theta |1\rangle - \sin\theta |2\rangle$,   
and $|\psi_{3}(\theta)\rangle = \frac{1}{\sqrt{2}} 
\big(\sin\theta |1\rangle + \cos\theta |2\rangle -|3\rangle \big)$.
Note that the eigenvectors are independent of $\rho$ and that none of 
them changes sign around any loop that does not pass through the origin. 
The function $F$ along the loop is
\begin{equation}\label{FTxt}
F(\theta) = \left( \begin{array}{ccc}
\frac{1}{\sqrt{2}}\sin\theta& \cos\theta  &\frac{1}{\sqrt{2}}\sin\theta\\
\frac{1}{\sqrt{2}}\cos\theta& -\sin\theta &\frac{1}{\sqrt{2}}\cos\theta\\
    \frac{1}{\sqrt{2}}        &       0       &     - \frac{1}{\sqrt{2}}\\
\end{array}\right).
\end{equation}
We aim to determine which loops in the $X$-$Y$ plane that map to
nontrivial loops in SO($3$) under $F$. The rotation angle $\phi$ 
and rotation vector ${\bf \hat{v}}$ of $F(\theta)$ are given by
\begin{eqnarray}
\phi(\theta) &=& 
\arccos \left(-1+\frac{1}{2}\left(1-\frac{1}{\sqrt{2}} 
\right)(1-\sin \theta)\right),\nonumber\\
{\bf \hat{v}}(\theta) &=& \frac{1}{N}(-\cos \theta,\sin
\theta-1,(1-\sqrt{2})\cos \theta) \label{v},
\end{eqnarray}
respectively, where $N$ is a normalization constant. Eq. (\ref{v})
shows that $\phi (\theta) = \pi$ only for $\theta = \pi/2$. It is also
clear that ${\bf
\hat{v}}$ is continuous everywhere except on the ray $\theta = \pi/2$,
where it abruptly changes sign. A plot of $\phi$ and ${\bf \hat{v}}$
as functions of $\theta$ appears in Fig. \ref{components}. Thus,
exactly the loops in the $X$-$Y$ plane that cross the ray $\theta
= \pi/2$ an odd number of times are mapped to nontrivial loops in
SO($3$). This is a concrete manifestation of the validity of the 
test, since any loop that passes $\theta = \pi/2$ an odd number of 
times must encircle the degenerate subset of ${\cal Q}$.

Consider now the case $n=4$. Let $F(t) = (F_{ij}(t))$ be a loop in
SO($4$) parametrized by $t \in {\bf R}$. The first column 
$f(t) \equiv (F_{11}(t),F_{21}(t),F_{31}(t),F_{41}(t))$ of $F(t)$ 
then represents an element of $S^3$ tracing out a loop. Define the 
matrix
\begin{eqnarray} 
\label{tr}
 & & T(f(t))
\nonumber \\ 
 & & = 
\left( \begin{array}{cccc} F_{11}(t)& F_{21}(t)& F_{31}(t)& F_{41}(t)\\
-F_{21}(t)& F_{11}(t)& -F_{41}(t)& F_{31}(t)\\
-F_{31}(t)& F_{41}(t)& F_{11}(t)& -F_{21}(t)\\ 
-F_{41}(t)& -F_{31}(t)& F_{21}(t)& F_{11}(t)  \end{array} \right) ,  
\end{eqnarray}
being orthogonal and continuous as a function of $f \in S^{3}$. It
follows that $T(t)\equiv T(f(t)) \in {\textrm{SO}}(4)$ since it can be
continuously connected to the identity matrix, corresponding to $f =
(1,0,0,0) = {\bf \hat{e}}_{1} \in S^{3}$. Furthermore, since $f(t)$ is
a loop in $S^3$, $T(t)$ is a trivial loop. We consider the
loop in SO($4$) given by $T(t)F(t)$. This loop is nontrivial exactly
when $F(t)$ is, since $T(t)$ is trivial. However, the loop $T(t)F(t)$
is simpler than $F(t)$. Carrying out the matrix multiplication, yields
\begin{equation}\label{nyloop}
T(t)F(t) = \left( \begin{array}{cccc} 1 & 0 & 0 & 0 \\
0 &A_{11}(t) &A_{12}(t) &A_{13}(t)\\
0 &A_{21}(t) &A_{22}(t) &A_{23}(t)\\ 
0 &A_{31}(t) &A_{32}(t) &A_{33}(t)\\  \end{array} \right). 
\end{equation}
Thus, the loop $T(t)F(t)$ is trivial if and only if the loop
$A(t) = (A_{ij}(t))$ in SO($3$) is trivial.

\begin{figure}[tbp]
\centering
\includegraphics[width=8cm]{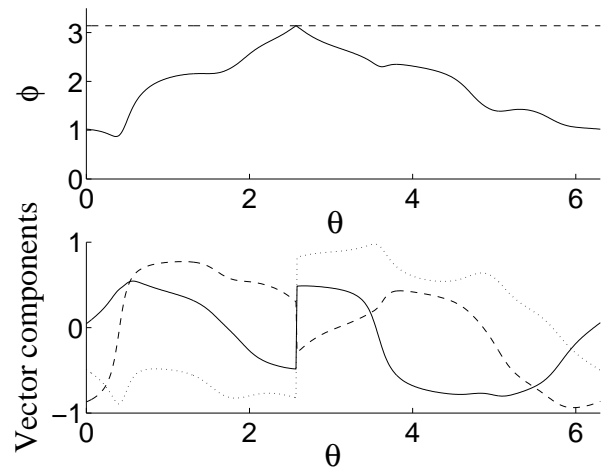}
\caption{The same plot as in Fig. \ref{components} but $\phi$ 
and the vector components are calculated from the matrix $A(\theta)$ 
of the $G\otimes g$ system along the loop $\Gamma$. Since $\phi$ 
equals $\pi$ at one point, $\Gamma$ must encircle a degeneracy.}
\label{components2}
\end{figure}

We have applied the above procedure numerically to the $G\otimes g$
Jahn-Teller system. The coupling matrix for this system reads 
\cite{OBrien94}
\begin{eqnarray}
\label{Gxg}
 & & H(g) = -qk^{G}_{g}\sqrt{2} 
\nonumber \\ 
 & & \times \left( \begin{array}{cccc}
g_{3}       &   g_{4}    & g_{1}-g_{3} &  g_{2}+g_{4} \\
g_{4}       &  -g_{3}    & -g_{2}+g_{4}  &  g_{1}+g_{3} \\
g_{1}-g_{3} & -g_{2}+g_{4} &   -g_{1}     &    g_{2}     \\
g_{2}+g_{4} &g_{1}+g_{3} &   g_{2}     &    g_{1}
\end{array} \right), 
\end{eqnarray}
where $g = (qg_{1}, \ldots, qg_{4})$ are normal modes and $k^{G}_{g}$
is the linear coupling constant. We consider a loop $\Gamma$ described
by $qg_{1} = qg_{4} = \cos \theta$ and $qg_{2} = qg_{3} = \sin \theta$, 
$\theta \in [0,2\pi]$. This loop encircles a four-fold degeneracy at the 
origin of parameter space and it can be checked that none of the 
eigenvectors changes sign along $\Gamma$. Fig. \ref{components2} shows the 
angle and vector of rotation of the matrix $A(\theta)$ along $\Gamma$. 
$\phi$ equals $\pi$ once, meaning that the loop in SO($4$) traced out 
by the eigenvectors is nontrivial. The four-fold degeneracy at the 
origin is thus detected.

Let us now consider whether there can be a better topological test
than the present one. It is easy to find loops encircling
degeneracies that map to trivial loops in SO($n$). This means that the
test does not find every degeneracy, so there may be room
for improvement. The rest of this Letter we devote to an argument
showing that if we are only allowed to consider eigenvectors on a loop
in parameter space ${\cal Q}$, and if the only information we have
about the Hamiltonian is that it is continuous, then there is no test
that can do better in implying degeneracies. To prove
this, assume that the eigenvectors $\{|\psi_{i}(Q(t))\rangle\}_{i=1}^{n}$ 
and corresponding eigenvalues $\{\lambda_{i}(Q(t))\}_{i=1}^{n}$ are 
known along the loop $\Gamma$ parametrized by $t\in {\bf R}$. Suppose 
that the eigenvectors are nondegenerate along $\Gamma$ so that
$\lambda_{1}(Q(t))<\lambda_{2}(Q(t))<\ldots<\lambda_{n}(Q(t))$ holds
for all $t$, and that the corresponding loop $F(Q(t))$ in SO($n$) is
trivial. Under these conditions we show that there always exists a
Hamiltonian $H(Q)$, continuous and nondegenerate on all ${\cal
Q}$, having exactly the eigenvectors $\{|\psi_{i}(Q(t))\rangle 
\}_{i=1}^{n}$ and eigenvalues $\{\lambda_{i}(Q(t))\}_{i=1}^{n}$ 
on $\Gamma$. Thus, if $F(Q(t))$ is trivial, the Hamiltonian can 
be nondegenerate everywhere, meaning that no test can imply a 
degeneracy.

For the proof we need to assume that the loop $\Gamma$ is homeomorphic
to $S^{1}$ and that there is a homeomorphism $D$ from ${\bf R}^{d}$
onto itself, such that $D(\Gamma)$ is the unit circle in the
$X_{1}$-$X_{2}$ plane. $(X_{1}, \ldots,X_{d}) = X = D(Q)$ denotes the
coordinates of the image of a point $Q \in {\bf R}^{d}$ under
$D$. Such a $D$ exists for all physically interesting $\Gamma$. 

We consider first the unit disc in the $X_{1}$-$X_{2}$ plane,
parametrized by the usual polar coordinates $\rho \in [0,1]$ and
$\theta \in [0,2\pi]$. By the homeomorphism $D$ we may view $F(Q(t))$
and $\{ \lambda_{i}(Q(t)) \}_{i=1}^{n}$ as functions of $\theta$,
i.e., $F(\theta) \equiv F \big( D^{-1} (\cos \theta, \sin \theta, 0,
\ldots, 0) \big)$ and $\lambda_{i}(\theta) \equiv \lambda_{i} \big( 
D^{-1}(\cos \theta, \sin \theta, 0, \ldots, 0) \big)$. The eigenvalues
can be continuously extended to the unit disc preserving the
nondegeneracy by defining $\tilde{\lambda}_{i}(\rho, \theta) = 
\rho \lambda_{i}(\theta) + (1-\rho)A_{i}$, where $A_{i}$ are any 
real numbers satisfying $A_{1}<A_{2}<\ldots<A_{n}$. Let 
$\tilde{\Lambda}(\rho, \theta) = \mbox{diag}[\tilde{\lambda}_{1}(\rho,
\theta), \ldots, \tilde{\lambda}_{n}(\rho, \theta)]$. Also the 
eigenvectors can be continuously extended to the disc. This
is exactly because they represent a trivial loop in SO($n$). 
Let the function $G:[0,1] \times [0,1] \rightarrow
\mbox{SO}(n)$ be a homotopy between the loop $F(\theta)$ and a
constant element $R_{0}$ of SO($n$) such that $G(0,s) = R_{0}$ for all
$s\in [0,1]$ and $G(1,\theta/2\pi) = F(\theta)$. A continuous
extension of $F$ to the whole disc is $\tilde{F}(\rho, \theta) =
G(\rho, \theta/2\pi)$.

We are now in a position to construct a suitable Hamiltonian as a
function of the coordinate $X$. Let $\rho$ take any value in
$[0,+\infty)$. Define
\begin{equation}\label{extHam1}
\tilde{H}(X)=
\left\{ \begin{array}{cc} 
\tilde{F}(\rho,\theta) \tilde{\Lambda}(\rho,\theta) 
\tilde{F}(\rho,\theta)^{T},&\mbox{if }\rho \leq 1 \\[1em]
\tilde{F}(1,\theta) \tilde{\Lambda}(1,\theta) \tilde{F}(1,\theta)^{T},
&\mbox{if }\rho > 1, \end{array}\right.
\end{equation}
where $T$ denotes matrix transposition. Note that $\tilde{H}(X)$
defined in this way is independent of the coordinates $X_{3}, \ldots,
X_{d}$, that it is continuous and nondegenerate for all $X$, and that
it has $\{ \ket{\psi_i (\theta)} \}_{i=1}^n$ as eigenvectors and $\{
\lambda_i(\theta) \}_{i=1}^n$ as eigenvalues on the unit circle in the
$X_{1}$-$X_{2}$ plane. To obtain the desired Hamiltonian as a function
of $Q$, simply define $H(Q) = \tilde{H}(D(Q))$.

In conclusion, we have described a test for degeneracies of real
Hamiltonians based on the behavior of their eigenvectors on a loop in
parameter space. If one considers a complete set of eigenstates, one
may detect a degeneracy even if none of the corresponding vectors
changes sign. The test works for all real quantum systems with finite
dimensional Hilbert spaces, including quantum billiards and
Jahn-Teller systems, and could find explicit use in quantum chemistry
applied to computed eigenvectors. We show also that no other
topological test can do better in detecting degeneracies.
\vskip 0.3 cm 
We wish to thank Tobias Ekholm for discussions. The work by E.S. 
was financed by the Swedish Research Council. 


\begin{thebibliography}{99}
\bibitem{remark} Note that if ${\bf \hat{v}}$ does not change sign for 
$\phi = \pi$, this corresponds to a gentle touch of the boundary of 
the ball, and does therefore not count as a piercing.
\end{thebibliography}
\end{document}